\documentstyle[11pt]{article}

\newcommand{\beq}{\begin{equation}}
\newcommand{\beqar}{\begin{eqnarray}}
\newcommand{\eeq}[1]{\label{#1} \end{equation}}
\newcommand{\eeqar}[1]{\label{#1} \end{eqnarray}}

\begin{document}    
\pagestyle{myheadings}
\thispagestyle{empty}
\mbox{ } \\         
\mbox{ } \vspace{0.6in}\mbox{ }\\
\begin{center}      
\begin{Large}       
{\bf                
Microscopic Enhancement
of Heavy-Element Production}\\[3ex]
 
\end{Large}         
P. M\"{o}ller\footnote{
Permanent address:{ 
\it P. Moller Scientific Computing and Graphics, Inc.,\
P.\ O.\ Box 1440, Los Alamos, New Mexico 87544, USA}}
 and J. R. Nix\\    
{\it Theoretical Division,
Los Alamos National Laboratory, Los Alamos, NM 87545, USA}\\[3ex]
P. Armbruster, S. Hofmann, and G. M\"{u}nzenberg\\
{\it Gesellschaft f\"{u}r Schwerionenforschung,
Planckstrasse 1, D64291 Darmstadt, Germany}\\[3ex]
{April 4, 1997}\\[1ex]
\end{center}        
\begin{small}       
{\bf Abstract.}     
Realistic fusion barriers are calculated
in a macroscopic-microscopic model
for several         
soft-fusion heavy-ion reactions leading to heavy and superheavy elements.
The results obtained in such a realistic  picture
are very different from those obtained in a purely macroscopic
model. For reactions on $^{208}$Pb targets, shell effects in the
entrance channel result in fusion-barrier energies at the
touching point that are only a few MeV higher than
the ground state for  compound systems near $Z=110$.
The entrance-channel fragment-shell effects remain
far inside the touching point, almost to
configurations only  slightly
more elongated than the ground-state configuration,
where the fusion barrier has risen to about 10 MeV above the
ground-state energy. Calculated single-particle level diagrams
show that few level crossings occur until the peak
in the fusion barrier very close to the ground-state
shape is reached, which indicates that dissipation
is negligible until very late in the fusion process.
Whereas the  fission valley in a macroscopic picture
is several tens of MeV lower in energy than is
the fusion valley, we find in the macroscopic-microscopic
picture that the fission valley is only about 5 MeV lower
than the fusion valley for soft-fusion reactions leading
to compound systems near $Z=110$. These results show that
no significant  ``extra-extra-push''
energy is needed to bring the system inside the
fission saddle point and that the typical reaction energies
for maximum cross section in heavy-element synthesis correspond to
only a few MeV above the maximum in the fusion barrier.

\end{small}         
 
\section{Introduction}
Cross sections for complete fusion in heavy-ion reactions
vary predictably and smoothly from the lightest compound systems
up to compound systems with nucleon number about $A=200$.
Beyond $A=200$, the compound-nucleus cross sections drop
dramatically [1--3].
A multi-dimensional macroscopic dynamical model that
explores a sufficiently general deformation space to
allow two touching spheres in the entrance channel to
evolve towards a single spherical shape {\it and\/} towards fission configurations
provides an explanation of the mechanism behind the sudden
drop in cross sections for heavy compound-nucleus
formation [4--7].   
 
For light projectiles and targets leading to compound systems
below $A=200$, the fusion and fission valleys are roughly
equivalent and the fission saddle point is more elongated than two
touching spheres. As the nucleon number $A$
increases above 200 the fission saddle-point shape rapidly becomes
less elongated and recedes to a more compact shape than
the fusion touching configuration.
At the touching configuration for these heavy nuclei, the steepest slope
of the potential-energy surface is towards the fission valley {\it outside\/}
the fission saddle point. Thus, if two colliding ions are brought together
with just enough energy to reach the touching configuration, the
system will evolve towards the fission valley and reseparate without
forming a compound system. A {\it necessary} condition [4,5]
for forming         
a compound system is that the dynamical trajectory evolves
to shapes inside the fission saddle point, that is, to shapes
between the spherical or deformed ground-state shape and the fission
saddle point. In a dynamical macroscopic model this can only be
accomplished by bringing the colliding heavy ions together
with energies substantially above the energy of the touching
configuration. The energy has to be higher both because
the maximum of the static  fusion barrier is located somewhat
inside the touching point for heavy systems and because
in a multi-dimensional deformation space
the topography of the potential is such that the
trajectory is deflected towards points outside the fission
saddle point. For collisions leading to nuclei between the actinides
and the superheavy elements it is the dynamical deflection of
the trajectory that is the most important effect.
The extra energy above the maximum in the fusion barrier that
is required to prevent deflection of the fusion trajectory to
the fission valley is often referred  to as the ``extra-extra-push''
energy [8,9].       
The results of      
such macroscopic dynamical calculations indicate that
for complete fusion leading to heavy elements in the
region $Z=100$--110 the optimum reaction energy is
several tens of MeV above the energy of the touching
configuration.

Some reaction cross sections agree with  this anticipated dependence on
the reaction energy, but most reactions cannot be even
qualitatively understood in terms of such a macroscopic
model. Particularly notable exceptions have been
observed in the synthesis of the six elements
with proton number  
$Z=107$--112        
[10--15]            
in soft-fusion reactions on targets near $^{208}$Pb. In these
reactions the optimum
cross sections are obtained below the macroscopic one-dimensional
Bass model of the fusion barrier [16,17]. The
optimum cross sections are obtained  at energies 8--4 MeV below
the Bass barrier,   
corresponding to  14--10  MeV excitation energy in the compound system.
Thus, the cross sections for these reactions cannot be understood in terms of
existing macroscopic models.
 
The soft-fusion reaction has long been thought to enhance
heavy-element evaporation-residue cross sections primarily because
it leads to compound nuclei of low excitation energy,
which enhances de-excitation by neutron emission relative
to fission. Higher excitation energies would lead to higher
fission probabilities. However, the evaporation-residue cross
section is the product of the
cross section for compound-nucleus formation
and the probability for de-excitation by
neutron emission.   
One may therefore ask if  soft fusion {\it also\/}
enhances the cross section for  compound-nucleus formation.
Because of the low excitation energies in the entrance
channel, the large negative shell correction
associated with target nuclei near the  doubly magic $^{208}$Pb
should be almost fully manifested at touching and slightly inside
touching.           
 
Nuclei near $^{258}$Fm have
already provided important insight into  fragment
shell effects in symmetric fission and
fusion configurations [18--20].
At $^{258}$Fm fission becomes symmetric with a very narrow
mass distribution, the kinetic energy of the fragments is
about 35 MeV higher than in the asymmetric fission of $^{256}$Fm
and the spontaneous-fission half-life is 0.38 ms for $^{258}$Fm compared to
2.86 h for $^{256}$Fm. These features are  well understood
in terms of the macroscopic-microscopic model. Shell effects
associated with division into fragments near $^{132}$Sn
lower the fusion valley at touching by  about $-20$ MeV
in the most favorable case, relative to a macroscopic potential-energy
model. This fragment shell effect remains important
far inside the touching point and results in fission into
the fusion valley with very compact {\it cold\/}
fragments for several fissioning nuclei in the vicinity of $^{258}$Fm.
Calculated single-particle level diagrams and potential-energy
surfaces show that these fragment shell effects start
to become important already slightly outside the first saddle
in the potential-energy surface. We now show that shell effects
are also very important in the fusion entrance channel in
soft-fusion heavy-ion reactions, which usually involve {\it asymmetric\/}
projectile-target combinations.
 
\section{The soft-fusion entrance channel}
 
Our calculations here of single-particle level diagrams and potential
energies are based on the finite-range droplet model  in its 1992
version [FRDM (1992)]. In this macroscopic-microscopic model we
use a generalized droplet model with
a Yukawa-plus-exponential term for the nuclear energy as the macroscopic model
and a realistic, diffuse-surface
folded-Yukawa single-particle potential as the starting point
for calculating the microscopic term by use of Strutinsky's method.
Special care has been taken to formulate the model so that the same
energy is obtained for the touching-sphere configuration both
when this configuration is considered as one highly deformed nucleus and
when it is considered as two separate nuclei with macroscopic and microscopic
interactions. The requirement that the energy be the same in these
two cases has led to the incorporation of shape-dependent
Wigner and pairing energies in the model and to a shape-dependent
smoothing range in the Strutinsky shell-correction method [19].
Complete            
details can be found in Ref.~[21].
 
For the symmetric fusion of two $^{132}$Sn nuclei the
minimum-energy configuration for the two merging ions corresponds
to two intersecting spheres for
a substantial part of the trajectory from the touching point
to the single sphere [20]. In our study here of the asymmetric
soft-fusion reactions we therefore select intersecting spheres
as our fusion shape 
configurations.     
As the deformation coordinate we use $r$, the distance between the
centers of mass of the two intersecting spheres. These centers of mass
coincide with the centers of the spheres {\it only\/} for the touching
configuration.

In Figs.~1 and 2    
we show calculated  proton and neutron single-particle level
diagrams for this sequence of shapes in terms of the $r$
shape coordinate for the reaction
$^{70}$Zn~+~$^{208}$Pb~$\rightarrow$~$^{278}$112.
This represents the reaction employed to reach the heaviest nucleus
known thus far. In the proton single-particle level diagram
in Fig.~1 the magic-fragment gap
combination $28+82=110$ remains far inside the touching point, up
to about $r/R_0=1.15$. The quantity $R_0$ is the radius of the spherical
compound system.    
This is in excellent agreement with the results of
calculations related to the symmetric fission of nuclei near $^{258}$Fm into
symmetric spherical fragments near $^{132}$Sn.
In the neutron single-particle level diagram in Fig.~2 the
magic-fragment gap combination $40+126=166$ is less prominent.
However, since the level density is  low both above and
below this neutron number the resulting microscopic correction
will also be  low. This region of relatively low single-particle
level density also remains for a substantial distance inside the
touching point, up to about $r/R_0=1.15$.
 
To quantitatively study the effect of the persistent
magic-fragment gaps on the fusion barrier as the heavy ions
merge we have calculated the fusion barrier for intersecting spheres
for three reactions of interest, namely:\\
$^{50}$Ti~+~$^{208}$Pb~$\rightarrow$~$^{258}$Rf\hfill
$^{68}$Zn~+~$^{208}$Pb~$\rightarrow$~$^{276}$112\hfill
$^{70}$Zn~+~$^{208}$Pb~$\rightarrow$~$^{278}$112 \\
which are shown in Figs.~3--5, respectively. Just beyond
the peak in the fusion barrier at about $r/R_0=1.0$ we have
switched from the intersecting-sphere parameterization
to Nilsson's perturbed spheroid $\epsilon$ parameterization so that
we accurately obtain the energy of the ground state.
The calculated ground-state shapes are indicated in the
figures. For each reaction we also
show the touching configurations and one intersecting-sphere
configuration at $r/R_0=1.04$, just before the maximum in the fusion
barrier. The dotted line shows the calculated fission barrier,
for which  we considered only $\epsilon_2$ and
$\epsilon_4$ shape distortions. The effect of mass asymmetry
on the fission barrier
is expected to be small for the two heavier nuclei, but up to
 2 MeV for the $^{258}$Rf barrier at distortions larger than
about $r/R_0=1.3$ [22].
The fusion barrier in the macroscopic FRDM without any shell effects
is given by the short-dashed line.
The touching configuration in all three cases is indicated by a thin
vertical long-dashed line. For the reactions in Figs.\ 3 and 5
the arrow indicates the incident energy corresponding to the
maximum evaporation-residue cross section.
 
\section{Discussion}
 
Clearly, the inclusion of microscopic effects in the calculation
of the fusion barrier has major consequences. Whereas the incident
energy corresponding to a maximum 1n evaporation-residue cross section
is                  
far below a macroscopic barrier [16,17], it
is at or slightly above the maximum in
our  realistic  fusion barrier. The maximum
occurs at about $r/R_0=1.0$, where our overlapping-sphere configuration may
not be general enough for an accurate calculation. From
comparisons with multi-dimensional calculations in the Fm region we
conclude that  exploring a more general
multi-dimensional parameterization may in some
cases lower the maximum fusion barrier by 2 or 3 MeV.
Thus, the incident energy is slightly above the maximum of
a still more realistic fusion barrier.
 
It has been argued earlier that shell effects in the
soft-fusion entrance channel favor compound-nucleus
formation both because  they  lower the fusion barrier,
so that fusion is possible at lower
energies [5,7,23],  
and because the persistence of the fusion valley far inside
the touching point prevents deflection towards the fission
valley [18].        
 
It is also of interest to understand how some of the kinetic energy in
the soft-fusion entrance channel is dissipated into internal energy.
If the dissipation occurs early in the fusion process
then extra energy over and above the fusion barrier would
be needed to push the system inside the saddle point.
However, it was argued earlier [3,24]
that level diagrams for
fission/fusion of $^{264}$Fm showed few level crossings occurring
in the merging of two $^{132}$Sn nuclei between touching and
about $r/R_0=1.15$ and in analogy the situation in soft fusion on
$^{208}$Pb targets should be similar. In that case, little dissipation
would occur until just outside the inner fusion and fission saddle points,
which would again reduce significantly the need for extra energy in the
entrance channel to reach the compound-nucleus configuration inside
the fission saddle point.
 
We see here in Figs.~1 and 2 that the analogy postulated in
Ref. [3,24] is      
borne out. The first crossings near the
Fermi surface do not occur until about $r/R_0=1.15$ for protons.
For neutrons, although the $N=166$ gap disappears for larger values of $r$,
the low level density associated with this neutron number also persists
until about $r/R_0=1.15$.
However, the large majority
of level crossings occur at $r/R_0=1.0$ or even inside, that is,
right at the peak of the fusion and fission barrier, very close
to the ground-state shape. This means that the original fragment
``cluster'' or shell structure is preserved during most of the
fusion process. This preservation of the original incident fragment
structure is also qualitatively clear from the appearance of
the intersecting-sphere shape at $r/R_0=1.04$ in Figs.~3--5.
This preservation of fragment character is maintained
even further inside the touching point than argued
and very schematically indicated in Fig.~18 of
Ref.~[3].           
 
Our present macroscopic-microscopic calculations  provide a much improved
understanding of the soft-fusion process, relative to a macroscopic-only
multi-dimensional   
picture [4--7]      
and relative to the one-dimensional
Bass model [16,17], 
which are both clearly insufficient to provide an
interpretation of the soft-fusion process. A microscopic
picture results in completely different fusion barriers,
potential-landscape valley structures, and dissipation mechanisms.
In addition, the dissipation depends critically on the
evolution of the fusion trajectory
and on the system location on this trajectory. The major part of the
dissipation takes place very near ground-state shape configurations.
\newpage            
 
\section{Conclusions}
 
The results here are the first step
in a completely new and more complete picture of soft fusion,
a picture that has the potential of eventually
 providing a more quantitative model of the magnitude
and the location in energy of the 1n evaporation-residue cross section.
The results here already show that the maximum cross section occurs a few
MeV above a {\it realistic\/} fusion barrier.
Our results also show that the differences
in the magnitude  of the 1n evaporation-residue
cross section can be interpreted  in terms of
a microscopic dissipation process at the inner fusion/fission barrier
and the dynamics of the fusion trajectory, which, just as in the
macroscopic picture, is located higher in energy than is the
fission valley. However, the fusion entrance channel is only 5 MeV or
so higher than the fission valley in the initial
stage of fusion and the difference is even smaller later,
in contrast to several tens of MeV differences in
a macroscopic-only picture.
An emission of a pre-compound neutron inside a realistic fusion barrier
would cool the compound system to an excitation energy that is below
the fission barrier.
For further understanding of the soft-fusion evaporation-residue
cross section,      
multi-dimensional calculations of the fusion/fission landscape
are required.       
Dynamical           
studies, properly incorporating
a microscopic dissipation model, of the trajectory in this potential-energy landscape
should provide additional insight.
 
\bigskip            
 
This work was supported by the U.~S. Department of Energy.

\newpage            
\itemsep=-0.2in     
\begin{center}                                                                  
{\bf References}                                                                
\end{center}                                                                    
\newcounter{bona}                                                               
\begin{list}%
{\arabic{bona})}{\usecounter{bona}                                              
\setlength{\leftmargin}{0.5in}                                                  
\setlength{\rightmargin}{0.0in}                                                 
\setlength{\labelwidth}{0.3in}                                                  
\setlength{\labelsep}{0.15in}                                                   
\setlength{\itemsep}{0.0in}
}                                                                               
\item                                                                           
P.\ Armbruster, Ann.\ Rev.\ Nucl.\ Part.\ Sci.\ {\bf 35} (1985) 135.            
                                                                                
\item                                                                           
G.\ {M\"{u}nzenberg}, Rep.\ Prog.\ Phys.\ {\bf 51} (1988) 57.                   
                                                                                
\item                                                                           
P.\ Armbruster, J.\ Phys.\ Soc.\ Jpn.\ {\bf 58} Suppl.\ (1989) 232.             
                                                                                
\item                                                                           
A.\ J. Sierk and J.\ R.\ Nix, Proc.\ Third IAEA Symp.\ on the physics and       
  chemistry of fission, Rochester, New York, 1973, vol.\ II (IAEA, Vienna,      
  1974) 273.                                                                    
                                                                                
\item                                                                           
J.\ R.\ Nix and A.\ J.\ Sierk, Phys.\ Scr.\ {\bf 10A} (1974) 94.                
                                                                                
\item                                                                           
P.\ {M\"{o}ller} and J.\ R.\ Nix, Nucl.\ Phys.\ {\bf A272} (1976) 502.          
                                                                                
\item                                                                           
P.\ {M\"{o}ller} and J.\ R.\ Nix, Nucl.\ Phys.\ {\bf A281} (1977) 354.          
                                                                                
\item                                                                           
W.\ J.\ Swiatecki, Phys.\ Scr.\ {\bf 24} (1981) 113.                            
                                                                                
\item                                                                           
W.\ J.\ Swiatecki, Nucl.\ Phys.\ {\bf A376} (1982) 275.                         
                                                                                
\item                                                                           
G.\ {M\"{u}nzenberg}, S.\ Hofmann, F.\ P.\ He{\ss}berger, W.\ Reisdorf, K.-H.\  
  Schmidt, J.~R.~H.\ Schneider, P.\ Armbruster, C.-C.\ Sahm, and B.\ Thuma, Z.\ 
  Phys.\ {\bf A300} (1981)~7.                                                   
                                                                                
\item                                                                           
G.\ {M\"{u}nzenberg}, P.\ Armbruster, F.\ P.\ He{\ss}berger, S.\ Hofmann, K.\   
  Poppensieker, W.\ Reisdorf, J.\ R.\ H.\ Schneider, W.\ F.\ W.\ Schneider,     
  K.-H.\ Schmidt, C.-C.\ Sahm, and D.\ Vermeulen, Z.\ Phys.\ {\bf A309} (1982)  
  89.                                                                           
                                                                                
\item                                                                           
G.\ {M\"{u}nzenberg}, P.\ Armbruster, H.\ Folger, F.\ P. He{\ss}berger, S.\     
  Hofmann, J.\ Keller, K.\ Poppensieker, W.\ Reisdorf, K.-H.\ Schmidt, H.\ J.\  
  {Sch\"{o}tt}, M.\ E.\ Leino, and R.\ Hingmann, Z.\ Phys.\ {\bf A317} (1984)   
  235.                                                                          
                                                                                
\item                                                                           
S.\ Hofmann, N.\ Ninov, F.\ P.\ He{\ss}berger, P.\ Armbruster, H.\ Folger, G.\  
  {M\"{u}nzenberg}, H.\ J.\ Sch{\"{o}}tt, A.\ G.\ Popeko, A.\ V.\ Yeremin, A.\  
  N.\ Andreyev, S.\ Saro, R.\ Janik, and M.\ Leino, Z.\ Phys.\ {\bf A350}       
  (1995) 277.                                                                   
                                                                                
\item                                                                           
S.\ Hofmann, N.\ Ninov, F.\ P.\ He{\ss}berger, P.\ Armbruster, H.\ Folger, G.\  
  {M\"{u}nzenberg}, H.\ J.\ Sch{\"{o}}tt, A.\ G.\ Popeko, A.\ V.\ Yeremin, A.\  
  N.\ Andreyev, S.\ Saro, R.\ Janik, and M.\ Leino, Z.\ Phys.\ {\bf A350}       
  (1995) 281.                                                                   
                                                                                
\item                                                                           
S.\ Hofmann, N.\ Ninov, F.\ P.\ He{\ss}berger, P.\ Armbruster, H.\ Folger, G.\  
  {M\"{u}nzenberg}, H.\ J.\ Sch{\"{o}}tt, A.\ G.\ Popeko, A.\ V.\ Yeremin, S.\  
  Saro, R.\ Janik, and M.\ Leino, Z.\ Phys.\ {\bf A354} (1996) 229.             
                                                                                
\item                                                                           
R.\ Bass, Phys.\ Lett.\ {\bf 47B} (1973) 139.                                   
                                                                                
\item                                                                           
R.\ Bass, Nucl.\ Phys.\ {\bf A231} (1974) 45.                                   
                                                                                
\item                                                                           
P.\ M{\"{o}}ller, J.\ R.\ Nix, and W.\ J.\ Swiatecki, Nucl.\ Phys.\ {\bf A469}  
  (1987) 1.                                                                     
                                                                                
\item                                                                           
P.\ M{\"{o}}ller, J.\ R.\ Nix, and W.\ J.\ Swiatecki, Nucl.\ Phys.\ {\bf A492}  
  (1989) 349.                                                                   
                                                                                
\item                                                                           
P.\ M{\"{o}}ller and J.\ R.\ Nix, J.\ Phys.\ G: Nucl.\ Part.\ Phys.\ {\bf 20}   
  (1994) 1681.                                                                  
                                                                                
\item                                                                           
P.\ M{\"{o}}ller, J.\ R.\ Nix, W.\ D.\ Myers, and W.\ J.\ Swiatecki, {Atomic    
  Data Nucl.\ Data Tables} {\bf 59} (1995) 185.                                 
                                                                                
\item                                                                           
P.\ {M\"{o}ller} and J.\ R.\ Nix, Nucl.\ Phys.\ {\bf A229} (1974) 269.          
                                                                                
\item                                                                           
A.\ S{\u{a}}ndulescu, R.\ K.\ Gupta, W.\ Scheid, and W.\ Greiner, Phys.\ Lett.\ 
  {\bf 60B} (1976) 225.                                                         
                                                                                
\item                                                                           
P.\ Armbruster, Proc.\ Int.\ School-Seminar on Heavy Ion Physics, Dubna,        
  Russia, 1986, Joint Institute for Nuclear Research Report No.\ D7-87-68       
  (1987) 82.                                                                    
                                                                                
\end{list}                                                                      
\newpage            
\begin{center}      
{\Large {\bf Figure captions}}\\[4ex]
\end{center}        
\newcounter{bean}   
\begin{list}        
{\Roman{bean}}{\usecounter{bean}
\setlength{\leftmargin}{0.55in}
\setlength{\rightmargin}{0.0in}
\setlength{\labelwidth}{0.5in}
\setlength{\labelsep}{0.05in}
}                   
\item[Fig.\ 1 \hfill]
Proton single-particle level diagram for merging nuclei in
an asymmetric heavy-ion collision leading to the  heaviest
known nucleus.      
The asymmetric configuration  in the entrance channel
leads to a mixing of states with odd and even parity.
The magic-fragment gaps associated with the
initial entrance-channel configuration remain far inside the
touching point, to about $r/R_0=1.15$, somewhat outside the maximum
in the fusion barrier.
 
\item[Fig.\ 2 \hfill]
Neutron single-particle level diagram for merging nuclei in
an asymmetric heavy-ion collision leading to the  heaviest
known nucleus.      
The asymmetric configuration  in the entrance channel
leads to a mixing of states with  odd and even parity.
 
\item[Fig.\ 3 \hfill]
 
Total and macroscopic fusion barriers for
the soft-fusion     
reaction            
$^{50}$Ti~+~$^{208}$Pb~$\rightarrow$~$^{258}$Rf
and                 
fission barrier corresponding to spontaneous fission from the
ground state.

\item[Fig.\ 4 \hfill]
 
Total and macroscopic fusion barriers for
the soft-fusion     
 reaction           
$^{68}$Zn~+~$^{208}$Pb~$\rightarrow$~$^{276}$112
and                 
fission barrier corresponding to spontaneous fission from the
ground state.

\item[Fig.\ 5 \hfill]
 
Total and macroscopic fusion barriers for
the soft-fusion     
 reaction           
$^{70}$Zn~+~$^{208}$Pb~$\rightarrow$~$^{278}$112
and                 
fission barrier corresponding to spontaneous fission from the
ground state.       
 
\end{list}          
\end{document}